# Broadband Single-Shot THz Sampling Using Reflection Gratings

C. Wu[1], M. Buzzi[1], A. Cavalleri[1,2*]

[1] *Max Planck Institute for the Structure and Dynamics of Matter, Hamburg, Germany*
[2] *Department of Physics, Clarendon Laboratory, University of Oxford, Oxford OX1 3PU, United Kingdom*
*\* andrea.cavalleri@mpsd.mpg.de*

**Single-shot electro-optic sampling (EOS) is a powerful method enabling the measurement of weak terahertz signals that would otherwise require prohibitively long acquisition times. This is generally achieved by encoding the EOS time delay into a spatial, angular, or frequency coordinate. In general, this technique operates well up to 3 THz but becomes more challenging for larger bandwidths, due to dispersion and imaging imperfections. Here, we demonstrate a reliable angular-encoding single shot EOS implementation that reaches frequencies beyond 6 THz. Diffraction simulations are used to design the experimental setup and adapt this technique to commercial reflection gratings, removing the need for custom-built echelon mirrors. Furthermore, we show that, contrary to earlier reports, group delay dispersion from angular dispersion does not reduce the bandwidth of single-shot EOS.**

Terahertz time domain spectroscopy has become a powerful technique in physics, chemistry and material science, and is routinely applied to study the dynamics of carriers and excitons in semiconductors [1,2], Cooper pairs in superconductors [3,4], insulator to metal transitions [5,6], and to characterize non-equilibrium phases of matter [7,8].

Key to this technique is the phase sensitive sampling of a THz waveform, typically measured before and after interaction with the sample under study. One way to achieve such sampling relies on the electro-optic (Pockels) effect in non-centrosymmetric crystals [9], where the instantaneous electric field of the THz pulse is imprinted onto the polarization of an ultrashort optical gate pulse. This process is generally referred to as electro-optic sampling (EOS). Typically, the full THz transient is mapped by analyzing the gate polarization while scanning the gate-THz delay. This procedure relies on varying the position of a mechanical delay stage, limiting acquisition speeds. Experiments in which the signal is small or require higher dimensional data acquisition [10–13] are frequently difficult with this method alone.

Various approaches have been proposed to accelerate EOS [14], most of them based on encoding a continuously variable THz-gate delay into a spatial, angular, or frequency coordinate of the gate pulse [15–17]. After interaction with the entire THz pulse, the gate pulse is either imaged or spectrally resolved to yield single-shot measurements of the complete THz waveform.

Angular-encoding methods are particularly robust and have been routinely implemented in time-domain THz spectroscopy experiments [18,19]. Yet, angle-encoded single-shot THz detection has so far been used to sample THz transients with spectral weight content up to 3 THz [14,20].

Here, we demonstrate an experimental implementation of this method that achieves bandwidths beyond 6 THz maintaining up to an order-of-magnitude reduction in acquisition time, compared to step-scan EOS. The design presented here is attractive also because it is based on a commercially available reflection grating, replacing costly custom-machined echelon mirrors.

## 1. Experimental Setup and Results

### A. Principle of angular encoding

Figure 1 illustrates the principle of angular-encoding single shot detection. The gate pulse diffracts off a grating and, due to angular dispersion, acquires an intensity-front tilt $\gamma$ with an angle given by [21]:

$$\gamma = \tan^{-1} \omega \frac{d\beta(\omega)}{d\omega} \quad (1)$$

where $\beta(\omega)$ is the diffracted angle after the grating, expressed as a function of the light frequency $\omega$. The diffracted pulse can be regarded as an array of gate beamlets that, owing to the intensity-front tilt, are focused onto the EOS crystal in a temporal sequence, and sample the THz transient at different time delays. As shown in figure 1, after the focus (EOS crystal), a second lens, arranged to achieve an overall 4f imaging geometry, projects an image of the grating onto the detector. This configuration enables the reconstruction of the polarization state of each beamlet using a spatially resolving photodetector. This scheme is generally referred to as "angular encoding", because different arrival times are associated with different propagation angles.

The time window $\Delta t$ accessible in a single-shot is determined jointly by the pulse front tilt angle introduced by the grating ($\frac{d\beta}{d\omega}$) and the size of the gate beam ($\Delta X$), and can be expressed as:

$$\Delta t = \frac{\Delta X \tan \gamma}{c} \cong \Delta X \cdot k_0 \frac{\partial \beta}{\partial \omega} \quad (2)$$

where $k_0$ is the wavevector of the center frequency.

The angular dispersion produces a spatial chirp at the focal plane of the first lens, with a full-width at half-maximum (FWHM) spread in the transverse (vertical in Fig. 1) direction given by:

$$\Delta x \cong f \frac{d\beta}{d\omega} \Delta \omega \quad (3)$$

where $\Delta \omega$ is the FWHM bandwidth of the gate pulse, and $f$ is the focal length of the first lens. A more detailed analysis of the gate pulse propagation shows that the front-tilted gate pulse acquires a negative group delay dispersion (GDD) proportional to propagation distance, which effectively stretches the pulse [21].

One question not well understood to date is if the negative GDD limits the temporal resolution of the single-shot detection geometry. In the following sections, after describing our implementation of a high-frequency single-shot THz detection setup and discussing its performance, we demonstrate—through wave-optics simulations—that under optimal detection conditions the temporal resolution is solely set by the intrinsic duration of the gate pulse and is not deteriorated by the negative GDD, a surprising result that enables the high performance of our apparatus.

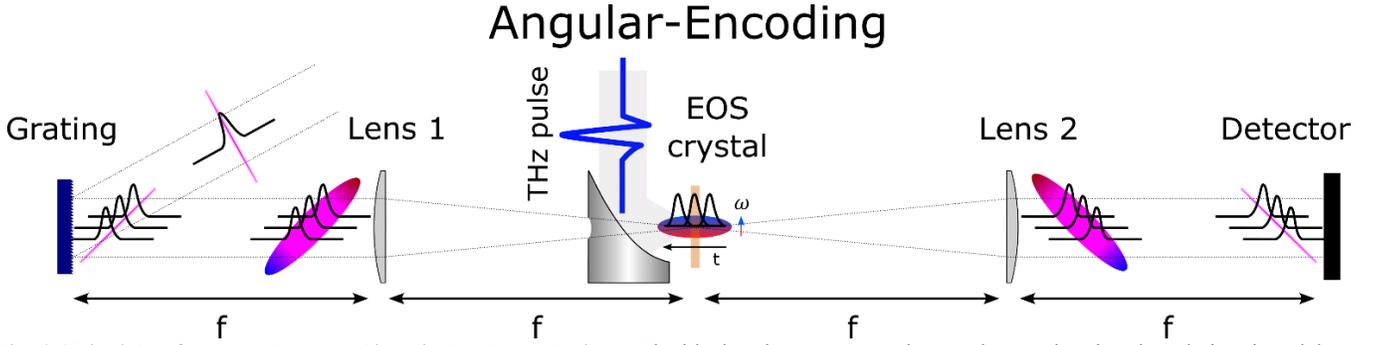

**Fig. 1. Principle of the angular-encoding single-shot technique.** The black pulses represent the gate beam, showing the relative time delays between the beamlets after the grating. The 4f geometry images the grating surface on the detector. The blue-to-red gradient represents the detailed spatial-temporal evolution of the gate pulse. Magenta color indicates the part of the pulse that contains all frequencies in the gate bandwidth. The THz pulse to be sampled is focused by a parabolic mirror and overlapped with the gate focus. Note that the gate focus is spatially chirped in the transverse direction (gradient-colored arrow) as a result of the angular dispersion caused by the grating.

### B. Experimental Realization and Results

Our experimental realization of angular encoded single-shot THz detection is shown in Fig. 2(a). A commercial Ti:Al$_2$O$_3$ amplifier operating at 2 kHz was split into a gate branch and a THz generation branch, modulated with a 1 kHz chopper. Broadband THz pulses were generated by optical rectification in a 200-μm-thick (110) GaP crystal. The residual 800 nm light was filtered using a Si plate placed at Brewster's angle to maximize THz transmission. The THz pulses were then focused with a parabolic mirror (f = 76.2 mm) onto a second 200-μm-thick (110) GaP crystal serving as the EOS crystal. To suppress interference from internal reflections the EOS crystal was optically contacted to a 1-mm-thick (100) GaP substrate for which the Pockels effect vanishes for transverse electric fields.

The gate beam (~7 mm 1/e$^2$ diameter) was expanded by a Keplerian telescope (6 times magnification) to illuminate the grating with a quasi-uniform intensity profile. A commercial 100-lines/mm diffraction grating (Richardson Gratings 53-*-011R) was used to introduce a pulse front tilt. The first diffraction order was directed to a plano-convex lens (f = 100 mm), focusing the gate beam onto the EOS crystal after being collinearly combined with the THz pulse using a pellicle beamsplitter (Thorlabs BP108). A second plano-convex lens (f = 125 mm) was placed a focal length away from the EOS crystal and was used to create an image of the grating on the spatially resolving balanced photodetector, implemented similarly to what reported in [17]. Changes in the gate polarization introduced by the electro-optic effect were measured using a quarter-wave plate and a Wollaston prism, which separated the orthogonal polarizations and directed them to the two CCD sensors of a dual line camera (Synertronic Designs Glaz LineScan-I-Gen2). To optimize signal collection, a set of cylindrical lenses focused the gate beam along the axis orthogonal to the angular-encoding direction, leaving the encoding unaffected. This setup enabled single-shot sampling over a ~3.2 ps time window, corresponding to a frequency resolution of ~0.3 THz.

A typical spectrum obtained with the single-shot THz detection scheme is shown in Fig. 2(b). The spectral content spans ~0.5–7 THz, with the high-frequency cutoff arising from an infrared-active phonon in GaP that constrains both the generation and detection bandwidth [22,23].

To benchmark the single-shot performance of our device, a reference spectrum was recorded using the conventional step-scanning method. In this measurement, the grating was replaced by a flat mirror, and the beam exiting the EOS crystal was directed to a balanced photodiode. The black solid line in Fig. 2(b) confirms that reliable single-shot detection is feasible even at 7 THz frequency.

These two measurements also allow us to quantify the gain in acquisition speed that single-shot detection provides compared to the traditional step-scan method. For each frequency in the spectrum, we defined a speed-up factor calculated as the ratio between the amounts of time required by both techniques to reach the same signal to noise ratio (see Supplemental Document S2 for detail). Fig. 2(c) shows the result of this analysis. For a wide frequency range spanning from 1 THz to 6.5 THz, the speed-up is at least 20 times, a value comparable to what reached by other setups optimized for lower frequency ranges.

while angular dispersion produces a spatially chirped gate, with a footprint on the EOS crystal that is more elongated when longer focal length lenses are used (see Eq. (3)). In the configuration of Fig. 3(a), the resulting spot-size mismatch between the tightly focused high-frequency THz components and the elongated gate causes only a part of the gate bandwidth to sample the THz pulse, effectively increasing the gating pulse duration, resulting in attenuation at high frequencies. On the other hand, low THz frequencies are unaffected irrespective of the chosen focal length, as their larger spot size still allows a good overlap with the gate. This effect is mitigated by the choice of shorter focal lengths as shown in Fig. 3(c).

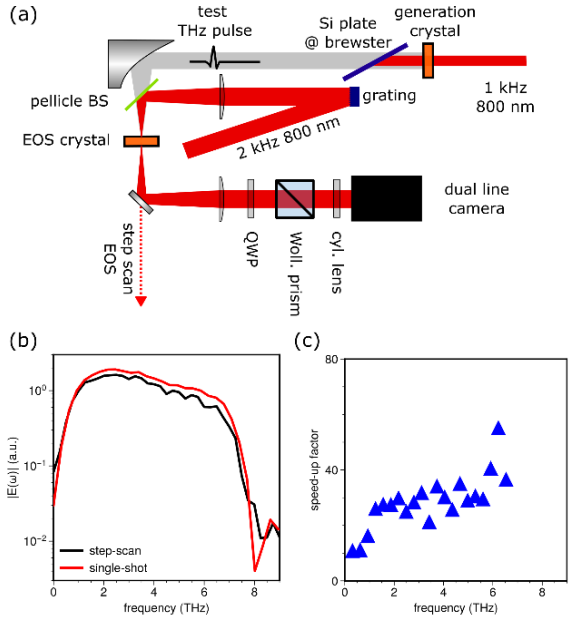

**Fig. 2. Setup and Results.** (a) Schematic of the single-shot THz detection setup. BS: beam splitter; QWP: quarter waveplate; Woll. prism: Wollaston prism; cyl. lens: cylindrical lens. (b) Comparison of the spectra obtained with the single-shot technique and with the traditional step-scanning technique. The two spectra are normalized to their respective signal at 1 THz. (c) Speed-up factor achieved by single-shot detection, calculated as the ratio of the time required by the two techniques to reach the same signal-to-noise ratio.

Next, we examined how the choice of the first focusing element influences the acquired spectra. Figures 3(a) and 3(c) show the two configurations considered. In the first case, a longer focal length lens (f = 200 mm) is used to focus the gate on the EOS crystal, while in the second configuration—the same as in Fig. 2—a shorter focal length lens (f = 100 mm) is employed. The corresponding THz spectra are presented in Figs. 3(b) and 3(d). The low-frequency components remain largely unaffected by the choice of lens, whereas the high-frequency content is strongly suppressed when using the longer focal length configuration (Fig. 3(a)). This behavior can be understood by comparing the gate-THz overlap at the focus in the two configurations (Fig. 3(a,c), lower panels). Diffraction effects cause higher THz frequencies to be focused to smaller spot sizes than lower ones (black circles),

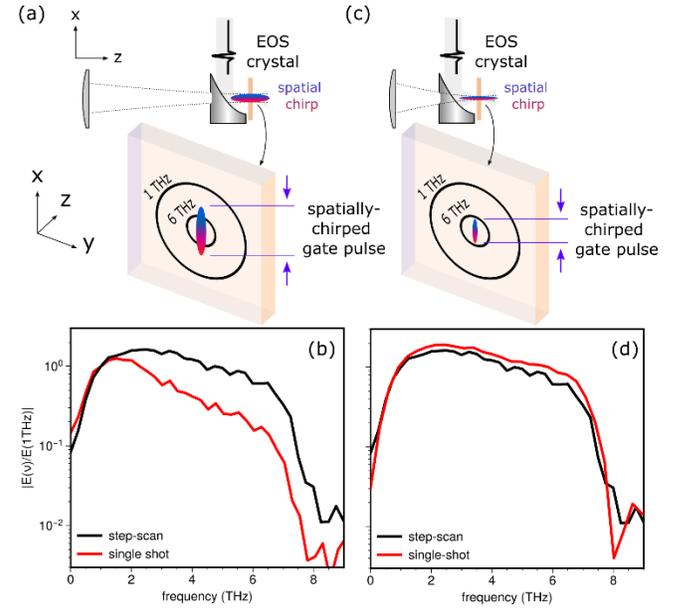

**Fig. 3. Comparison of two realizations of single-shot THz detection.** All THz spectra are normalized to 1 THz. (a) The gate is focused by a 200 mm lens and the spatially chirped gate focus spot elongates beyond the high frequency THz foci, as light of shorter wavelengths is focused to a smaller spot. Those THz frequencies overlap with only a narrower gate bandwidth, and thus are effectively sampled by a longer gating pulse. (b) At high frequencies, due to the effective longer sampling pulse duration, the single-shot response is suppressed compared to the step-scan one. (c) Same as (a) but the gate is focused by a 100 mm lens, resulting in a spot size smaller than the 6 THz spot. The THz spectrum measured with single shot now matches that obtained with the step-scan technique, as shown in (d).

## 2. Bandwidth of the device

To better understand the factors limiting the time resolution of a single-shot detection setup—and thus the highest frequency that can be faithfully sampled—we simulated a simplified experiment using wave-optics calculations as illustrated in Fig. 4.

In this thought experiment, the time-resolved detection scheme is identical to that used in single-shot THz measurements, except that only intensity changes are monitored at the detector. A sample was placed at the position of the EOS crystal, and we assumed that a pump pulse abruptly switched the crystal from transparent to opaque, as shown in the lower left panel of Fig. 4. In this configuration, the detector measures the single-shot time resolved transmission curve of the sample.

Using wave-optics methods (see Supplemental Document S1), a 35-fs gate pulse was propagated from the grating to the sample at the focal plane of the first lens, modulated by the sample's time-dependent transmission function, and then propagated to the detector to simulate the acquired trace. The simulation results are shown in the lower-right panel of Fig. 4. In the simulated single-shot trace, the sudden temporal change in the sample's transmission produces a gradual spatial variation in the intensity. Fitting this trace with an error function yielded the temporal resolution of the sampling, showing that it is determined solely by the 35-fs gate pulse duration and is unaffected by the negative group delay dispersion introduced by the grating. This result indicates that the beamlets maintain their original pulse duration and are not affected by the negative GDD introduced by the grating discussed in [21] that rather affects the pulse intensity front as a whole as shown in Supplemental Document S1.

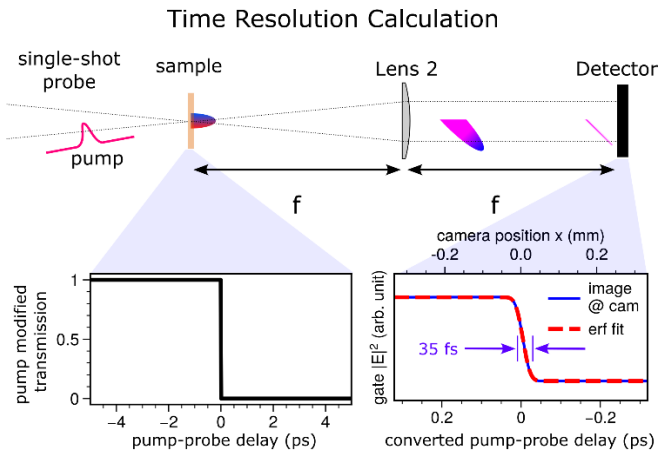

**Fig. 4. Calculation of the time-resolution.** A pump pulse strikes a sample placed in the gate focus at $t = 0$ and abruptly turns the initially transparent sample opaque to the probe, as shown in the lower left panel. The spatio-temporal evolution of the gate pulse is simulated from the grating to the sample and to the detector. Due to the time dependent transmission change, the second half the gate beamlet train is blocked. After propagation to the detector, the intensity profile is extracted and an error function fit is used to extract the time resolution yielding a value equal to the initial pulse duration of the probe pulse. In other words, the temporal resolution is not affected by angular dispersion.

While this estimate holds for beamlets propagating through the lens center, beamlets traveling away from the center require an additional consideration. As illustrated in the top panel of Fig. 5, beamlets from the lens edge reach the sample at an angle, overlapping with the pump pulse non-collinearly. As in conventional non-collinear pump-probe experiments, this leads to a smearing of the time resolution [24,25]. Consequently, in angle-encoding single-shot setups the experimental time resolution varies throughout the time window, since beamlets arriving at different times traverse different regions of the lens. This effect was briefly noted in [26] and a simple approximate analysis is provided in Supplemental Document S2. Here, we explicitly calculated the variation of the temporal resolution throughout the time window by repeating the simulation of Fig. 4 while varying the pump arrival time. The results are shown in the lower panels of Fig. 5 for two gate pulse durations of 100 fs and 35 fs. Repeating the analysis of Fig. 4, we extracted the experimental time resolution at different points in the time window. For 100-fs gate pulses, the time resolution broadens only slightly up, to ~105 fs, whereas for 35-fs pulses the broadening is more pronounced, reaching ~80 fs. The dependence of the broadening on the initial pulse duration arises because shorter duration beamlets, carry more bandwidth and, owing to angular dispersion, have a larger footprint on the sample, leading to a larger temporal smearing (Fig. 5 lower right panels).

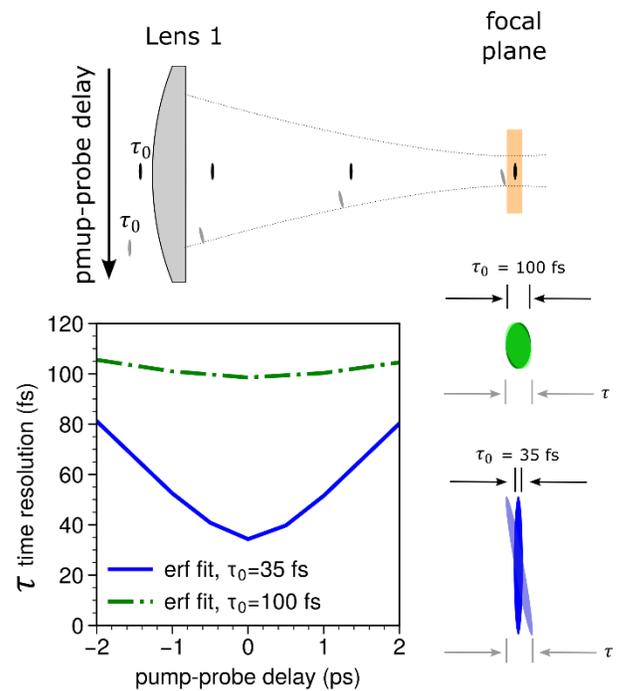

**Fig. 5. Time-resolution distortions caused by the focusing lens.** A beamlet going through the lens away from the optical axis would overlap with the pump non-collinearly. This effectively worsens the time resolution, which is calculated explicitly by the same approach as in Fig. 4. The blue solid curve and the green dash-dotted curve represent the results for pulses with initial (before grating) pulse duration of 35 fs and 100 fs, respectively. The lower right panel illustrates the amount of broadening for these two different initial pulse durations.

In summary, we described an experimental implementation of an angular-encoded single-shot THz detection scheme that produces a significant speed-up in acquisition time up to

7 THz frequency, compared to step-scan detection. Through diffraction simulations, we showed that the sampling duration is not affected by the chirp introduced by the grating. Instead, the quality of gate-THz overlap and the noncollinear gate-THz angle concomitant with the angular-encoding are the principal cause for the degradation of the temporal resolution.

## Back Matter

**Acknowledgment.**

The authors acknowledge insightful discussions with A. Liu and C. Trallero.

**Disclosures**.

The authors declare no conflicts of interest.

**Data Availability Statement (DAS).**

Data underlying the results presented in this paper is available from the corresponding author upon request.

**Supplemental Document.** See Supplemental Document for supporting content.

## References.

**Supplementary Material for "Broadband Single-Shot THz sampling using reflection gratings" (C. Wu, et al.)**

## S1. Diffraction calculations: propagation of a front-tilted beam

The propagation of the gate beam through the different components of the single-shot detection setup was simulated in the frequency domain, restricting the calculation to a single spatial dimension corresponding to the frequency dispersion direction of the grating.

To model the evolution of the gate pulse through the different elements of the single-shot setup we considered their effect and the free space propagation between them. This was done as described below for each frequency component in a range that covers the entire gate pulse bandwidth. These components were then summed and Fourier transformed back in the time domain to obtain the gate pulse time-dependent electric field at different location of the single-shot detection setup.

Free space propagation from a plane at $z$, to one at $z'$ is accounted for using the angular spectrum algorithm [1]. We first introduce the angular spectrum as:

$$A\left(\frac{\alpha}{\lambda};z\right) = \int E(x,z) \exp\left[-i\frac{2\pi}{\lambda}(\alpha x)\right] dx$$

where $E(x;z)$ is the electric field profile at the initial plane, and $\alpha$ is the directional wavevector cosine along $x$. Due to propagation, the phase added to a specific wave component is $\exp\left[ik(z'-z)\sqrt{1-\alpha^2}\right]$, hence we have [2]:

$$E(x;z') = \int A\left(\frac{\alpha}{\lambda};z'\right) \exp\left[i\frac{2\pi}{\lambda}(\alpha x)\right] d\frac{\alpha}{\lambda}$$

on the final plane, where the propagated angular spectrum is given by:

$$A\left(\frac{\alpha}{\lambda};z'\right) = A\left(\frac{\alpha}{\lambda};z\right) \exp\left[ik(z'-z)\sqrt{1-\alpha^2}\right]$$

This is equivalent to the Rayleigh-Sommerfeld diffraction theory, without paraxial or far-field assumptions.

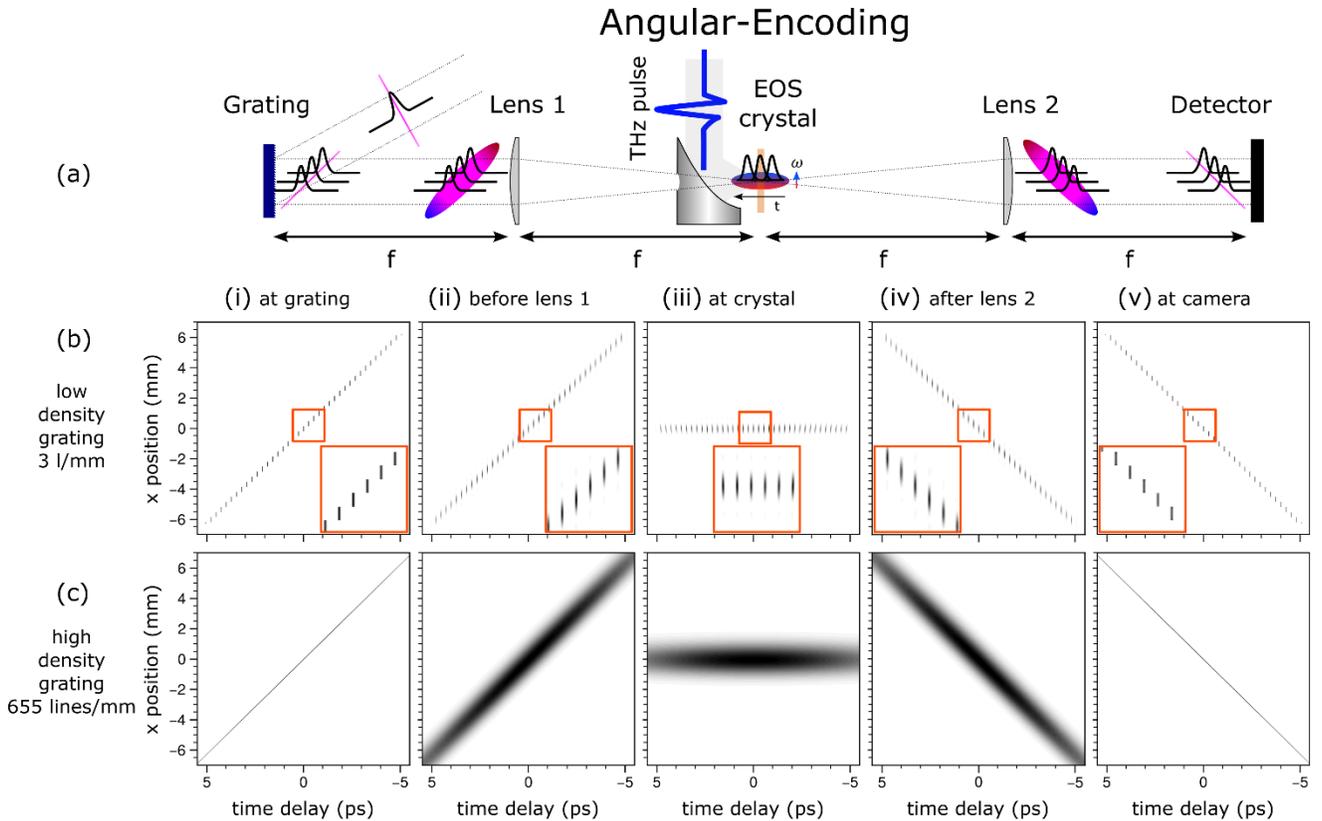

**Fig. S1.** (a) Schematic of the angular-encoding single-shot detection setup. (b, c) Results of the diffraction calculation. Two grating periods are chosen: one in the discrete limit (b) where the grating can be thought of as an array of mirrors and one in the continuous limit (c) where the grating pitch is comparable to the wavelength. A spatio-temporal map of the gate pulse is shown at five positions in the 4f system to illustrate the effect of each element.

The effect of the lenses and the grating was modeled directly in real space. Hence, for each frequency, the electric field right after the lens is defined as:

$$E_{after}(x) = E_{before}(x) \exp\left(-i\frac{k(\omega)}{2f}x^2\right)$$

Similarly, the effect of the grating was modeled as:

$$E_{after}(x) = E_{before}(x) \exp\left(-ik_0\frac{d\beta}{d\omega}(\omega-\omega_0)x\right)$$

where $k_0$ is the wavevector modulus of the center frequency, $\frac{d\beta}{d\omega}$ is angular dispersion with $\beta(\omega)$ being the diffracted angle after the grating as a function of the light frequency $\omega$, $\omega_0$ is the gate-pulse center frequency, and $x$ is the spatial coordinate along the grating surface. While for high groove density gratings the introduced phase was modeled as continuous along the spatial direction [3], for low groove density the additional phase followed a step-like function.

$$e^{-ik_0\frac{d\beta}{d\omega}(\omega-\omega_0)x} \rightarrow e^{-ik_0\frac{d\beta}{d\omega}(\omega-\omega_0)d\cdot\left\lfloor\frac{x}{d}\right\rfloor}$$

The discretization needed to perform these simulations was set up using the parameters reported in Table 1. The choice of the grid density allowed to fully capture diffraction process also from gratings with very high density.

Table 1. Typical parameters for simulation

| Parameter | value |
|---|---|
| Number of grid points in space $Nx$ | $2^{15} = 32768$ |
| Size of the grid in spatial coordinate $Lx$ | 50 [mm] |
| Number of grid points in frequency $Nw$ | $2^{12} = 4096$ |
| Center wavelength $\lambda_0$ | 800 [nm] |
| Light pulse FWHM bandwidth | 27 [nm] |
| Size of the grid in frequency domain | 20x FWHM |

Fig. S1 shows the space–time profile of the gate pulse at five representative positions: (i) immediately after the grating, where the pulse front is tilted, (ii) just before the first lens, (iii) at the focal plane, (iv) immediately after the second lens, and (v) at the detector. The simulation was repeated for two grating densities: a low density (upper row), representing a stepped mirror, and a high density (lower row), representing a typical diffraction grating.
In the low-density grating case, the grating pitch is large and each groove acts as a small mirror, producing an array of smaller time delayed beamlets that do not diffract significantly before entering the focusing lens. The pulse duration of each beamlet remains the same as the initial pulse duration, hence it is clear that the time resolution remains unaffected.
In the high-density grating case, the pitch is small and each groove introduces a significant diffraction effect. The pulse duration of the intensity front tilted gate pulse can be estimated by taking a horizontal line cut of the spatio-temporal map right before the focusing lens (position (ii), Fig. S1(c)). This obtained duration matches well with Eq. 23 in Ref. [3] and reported here (we set the parameter $u = 0$, neglecting effects due to the finite beam size):

$$\tau(z) = \tau_0\sqrt{1+\left(\frac{k_0\left(\frac{d\beta}{d\omega}\right)^2 z}{\tau_0^2}\right)^2}$$

where $z$ is the propagation direction, and $k_0$ is the wavevector of the center frequency. This effect is known as group velocity dispersion introduced by angular dispersion (GVD-AD) [3]. Here, it is less obvious to infer directly the duration of the gate pulse at the focus position (EOS crystal) and hence the sampling time resolution. However, additional calculations shown in Fig.4, show that it remains determined by the initial gate pulse duration.

## S2. Calculation of the speedup factor

The frequency dependent speed-up factor is defined as the ratio of the time cost required by the step-scan and single-shot methods to reach the same signal-to-noise ratio (SNR). The SNR is frequency dependent, and is defined as:

$$\text{SNR}(\nu) \equiv \frac{E(\nu)}{\bar{\sigma}(\nu)}$$

where $E(\nu)$ is the signal frequency dependent mean value and $\bar{\sigma}(\nu)$ is the frequency dependent standard error over repeated measurements.
Fig. S2 shows the frequency dependent SNR achieved by single-shot detection averaging an increasing number of gate laser shots ranging from 640 to 8000, and for step-scan measurements. The step-scan measurement was performed using 3 μm delay steps (i.e., 20 fs) and averaging 120 gate laser pulses per delay point. A time window of 3.2 ps was considered for both methods.
The total time cost for each measurement is reported in the legend of Fig. S2 and is calculated as follows. For single-shot detection, additional processing time adds ~12.5% to the averaging time (the time required to accumulate N shots at a given repetition rate). For step-scan, the time cost includes both the averaging time and that required for the mechanical delay stage to move and stabilize. For the parameters mentioned above the gate delay stage motion required to ~40 s, while averaging contributed 10 s.
As seen in Fig. S2, a single-shot measurement taking ~2 s achieves a SNR of ~200 at 2 THz, while step-scan requires a total of ~50 s to reach the same figure, giving rise to a single-shot speed-up factor of ~ 25 times, as stated in the main text.

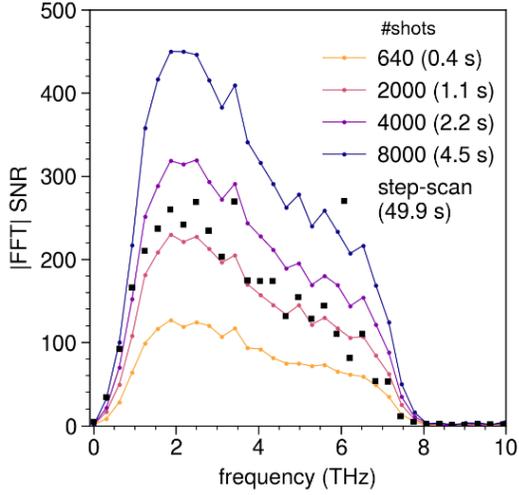

**Fig. S2.** Frequency dependent signal-to-noise ratio achieved with increasing number of averages with single-shot detection compared to that with the step-scan method.

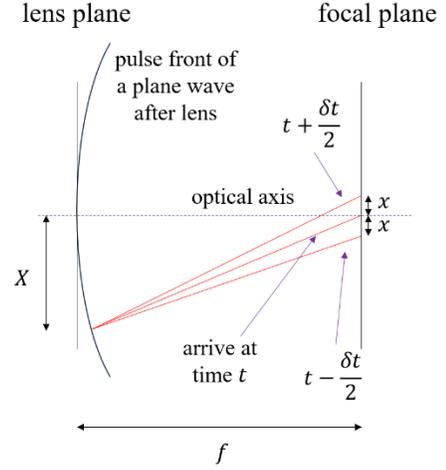

**Fig. S3.** Raytracing analysis for the distortions caused by the focusing lens.

## S3. Raytracing estimate for the time delay dependence of the temporal resolution

The results presented in Fig. 5 can also be obtained from a simple analytical raytracing modeling. Fig. S3 shows the raytracing path, for a point at a distance $X$ from the focusing lens optical axis, representing a sampling time $t = X \cdot k_0 \frac{\partial \beta}{\partial \omega}$. In this case the difference in arrival time at the focal plane (sample/EOS crystal position) of the reddest and bluest part of each single shot gate beamlet can be expressed as:

$$c\delta t = \sqrt{f^2 + (X+x)^2} - \sqrt{f^2 + (X-x)^2} \approx \frac{2X \cdot x}{\sqrt{f^2 + X^2}}$$

with $2x = f \frac{\partial \beta}{\partial \omega} \Delta\omega$ (FWHM spread of focus size of the angular dispersed gate) and $\Delta t = X \cdot k_0 \frac{\partial \beta}{\partial \omega}$ (space-to-time conversion, $\Delta t$ is the time delay), we obtain:

$$c\delta t \approx \frac{X}{f} \cdot f \frac{\partial \beta}{\partial \omega} \Delta\omega = X \frac{\partial \beta}{\partial \omega} \Delta\omega = \Delta t \frac{\Delta\omega}{k_0}$$

This is equivalent to the calculation reported in Ref. [5], which gives the asymptotic temporal resolution when the beamlet enters very far away from the center of the lens and the arrival time at the focus between the leading and trailing part of the angularly dispersed gate beamlet becomes comparable with the gate pulse duration itself.

Importantly, this represents the ultimate limit for angular-encoded single shot detection. If the gate pulse center frequency and bandwidth are fixed, no technical expedient can mitigate this effect.

## S4. Additional Experimental Data with a higher density grating

In the following we report additional results acquired with a single-shot detection setup similar to that shown in Fig. 2 but using 0.5 mm ZnTe (100) as generation crystal, a grating with 300 lines/mm groove density (Thorlabs GR25-0310) and a first-lens focal length of 200 mm.

In these conditions the pulse length right before the lens can be estimated as described in section S1. The introduced GDD-AD is 16321 fs², which would stretch the 35 fs gate pulse to 1.3 ps.

Fig. S4(a) shows the reference spectrum (black) for this setup acquired using the step-scan method. The green line is the spectrum obtained convolving the step-scan time trace with a 1.3 ps FWHM Gaussian function. All frequencies above ~0.5 THz are significantly attenuated, due to the longer gate pulse duration.

For comparison Fig. S4(b) shows the spectrum acquired in the same conditions using single-shot detection. The agreement with the reference spectrum (black line, Fig. S4(a)) is very good and the spectral bandwidth is the same, showing that single-shot sampling is not affected by the GDD introduced by the angular dispersion.

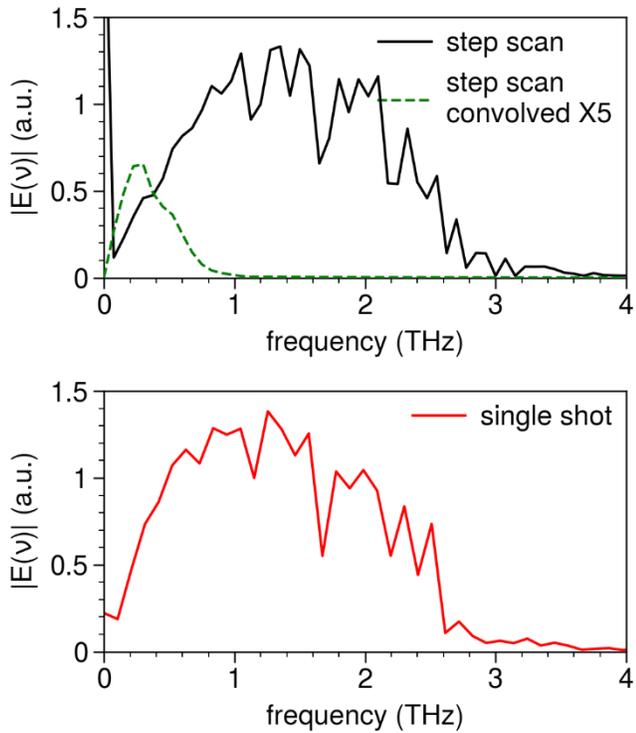

**Fig. S4. Comparison of EOS signal in frequency domain with ZnTe generation, GaP detection.** (Upper panel) The black curve is the reference step-scan result, and the green dashed curve is Fourier transform of the step-scan result convolved with a gaussian filter of FWHM 1.3 ps, multiplied manually by 5 for better comparison. (Lower panel) The red curve is the single shot result, which shows spectral weights up to 2.5 THz.

**References.**